\documentclass[10pt,a4paper,twoside]{article}
\usepackage{graphicx}
\usepackage{baltlat6}
\usepackage{array}
\usepackage{here}
\begin{document}
\ \
\vspace{0.5mm}
\setcounter{page}{277}
\vspace{8mm}

\titlehead{Baltic Astronomy, vol.\,?, 2011}

\titleb{FIR/SUBMM SPECTROSCOPY WITH HERSCHEL: \\FIRST RESULTS FROM THE VNGS AND
  H-ATLAS SURVEYS}

\begin{authorl}
\authorb{Maarten Baes}{1},
\authorb{Jacopo Fritz}{1},
\authorb{Naseem Rangwala}{2},
\authorb{Pasquale Panuzzo}{3},
\authorb{Christine D. Wilson}{4},
\authorb{Steve Eales}{5},
\authorb{Ivan Valtchanov}{6} and
\authorb{the VNGS and H-ATLAS consortia}{}
\end{authorl}

\begin{addressl}
\addressb{1}{Sterrenkundig Observatorium, Universiteit Gent,
  Krijgslaan 281 S9. B-9000 Gent, Belgium; maarten.baes@ugent.be}
\addressb{2}{University of Colorado, Boulder, USA}
\addressb{3}{CEA, Laboratoire AIM, Gif-sur-Yvette, France}
\addressb{4}{McMaster University, Hamilton, Canada}
\addressb{5}{Cardiff University, Cardiff, UK}
\addressb{6}{European Space Astronomy Centre, Villanueva de la Ca\~{n}ada, Spain}
\end{addressl}

\submitb{Received: \today; accepted: \today}

\begin{summary} 
  The FIR/submm window is one of the least-studied regions of the
  electromagnetic spectrum, yet this wavelength range is absolutely
  crucial for understanding the physical processes and properties of
  the ISM in galaxies. The advent of the Herschel Space Observatory
  has opened up the entire FIR/submm window for spectroscopic
  studies. We present the first FIR/submm spectroscopic results on
  both nearby and distant galaxies obtained in the frame of two
  Herschel key programmes: the Very Nearby Galaxies Survey and the
  Herschel ATLAS.
\end{summary}


%% \resthead is the RUNNING TITLE at top of the pages
\resthead{FIR/submm spectroscopy with Herschel}
{M. Baes et al.}

\sectionb{1}{INTRODUCTION}

The FIR/submm represents one of the least-studied regions of the
electromagnetic spectrum, yet this wavelength range is absolutely
crucial for understanding the physical processes and properties of the
ISM in galaxies (i.e.\ Tielens \& Hollenbach 1985). One of the most
fundamental issues in ISM studies is to address what factors control
the heating and cooling processes in the different phases. FIR/submm
spectroscopy is a unique tool to answer this question, as the
FIR/submm window contains important cooling lines of different phases
of the ISM. The fine-structure line of singly ionized carbon
[C{\sc{ii}}] at 158 $\mu$m is the most important cooling line of the
neutral ISM of normal starforming galaxies. Other important atomic
fine-structure lines include the [O{\sc{i}}] lines at 63 and 145
$\mu$m and the [C{\sc{i}}] lines at 370 and 609 $\mu$m. The FIR/submm
window also contains a number of atomic fine-structure lines that
trace ionized gas, such as the [N{\sc{ii}}] lines at 122 and 205
$\mu$m, and the [O{\sc{iii}}] lines at 52 and 88 $\mu$m. Finally, it
contains a large variety of high-$J$ rotational transitions of
different molecules, including CO, HCN, HCO$^+$, H$_2$O and many
others. The low-$J$ transitions (typically up to $J=4-3$) of many
molecules can be observed with ground-based radio and submm
telescopes, but they only trace the cooler gas and not the warmer gas
that can dominate the cooling of the molecular ISM. The mid- to
high-$J$ transitions, which are either difficult to observe or
completely inaccessible from the ground, have a large span in critical
densities, making them excellent tracers of the physical conditions of
gas over a wide range in temperatures and densities. They are also 
excellent probes to distinguish between different energy sources
responsible for the excitation of the gas, for example starbursts or
AGN (e.g.\ Spaans \& Meijerink 2008).

The main difficulty with FIR/submm spectroscopy is the fact that the
earth atmosphere is almost completely opaque to FIR/submm radiation,
such that these observations need to be performed from the
stratosphere or from space.  The first FIR spectroscopic observations
were done with KAO, COBE and balloons in the 1980 and early 1990s. A
major step forward was the launch of ISO with its LWS spectrograph,
which made studies of large samples of galaxies of different types
possible (e.g.\ Brauher et al.\ 2008). The wavelength region longwards
of 200~$\mu$m remained closed, however, until the advent in 2009 of
the Herschel Space Observatory (Pilbratt et al.\ 2010). Herschel has
three instruments onboard, which cover the FIR/submm wavelength region
with unprecedented sensitivity and spatial resolution. PACS
(Potglitsch et al.\ 2010) operates either as an imaging photometer or
an integral field spectrometer over the spectral band from 55 to
210~$\mu$m. SPIRE (Griffin et al.\ 2010) performs simultaneous imaging
in three bands centered at 250, 350 and 500~$\mu$m, and contains an
imaging Fourier Transform Spectrometer (FTS) which covers
simultaneously its whole operating range between 194 and
671~$\mu$m. Finally, HIFI (de Graauw et al.\ 2010) is a
high-resolution heterodyne spectrometer with a resolution up to 10$^7$
that can cover the wavelength region from 157 to 625~$\mu$m.

In this paper, we present a number of recent FIR/submm spectroscopy
results on both nearby and distant galaxies obtained in the frame of
two Herschel key programmes: the Very Nearby Galaxies Survey and the
Herschel ATLAS. It should be noted that these results are just the
first results, and that many more of them (both detailed studies on
individual objects and more statistical studies on larger samples) are
expected to come out in the next few years.

\sectionb{2}{SPECTROSCOPY OF NEARBY GALAXIES}

\subsectionb{2.1}{The Very Nearby Galaxies Survey}

The Very Nearby Galaxy Survey (VNGS) is a Herschel Guaranteed Time Key
Project focusing on twelve galaxies within 25 Mpc and the archetypal
starburst galaxy Arp\,220. The galaxies show a diverse range of masses
and properties, from low-mass, late-type galaxies such as NGC\,2403 to
the radio galaxy Cen\,A. For all targets, a wealth of ancillary data
at virtually all wavelengths is available. The design of the survey
incoporates photometry and spectrometry with both PACS and SPIRE. The
survey aims to gain a detailed understanding of the processes that
regulate the ISM, and how these processes vary with the environment
within different galaxies. The detailed study of these resolved
galaxies will not only act as a benchmark for studies of more distant
galaxies, but will also bridge the gap between surveys of distant
objects and the extensive Galactic surveys which have superior
physical resolution but by their nature are limited to observations of
one galaxy.

The first results from the VNGS were mainly photometric studies. These
include the determination of the nature of dust heating in M81, M83
and NGC\,2403 (Bendo et al.~2010, 2011), an attempt to separate
Galactic cirrus emission from extragalactic dust emission in M81
(Davies et al.~2010), a study of the dust distribution and heating
mechanisms in and around M82 (Roussel et al.~2010) and a detailed look
at extragalactic dust clouds around the radio galaxy Cen~A (Auld et
al.\ 2011). Here we focus on two spectroscopic studies with the SPIRE
FTS of two VNGS galaxies: the starburst galaxy M82 (Panuzzo et al.\
2010) and the ULIRG Arp\,220 (Rangwala et al.\ 2011).

\subsectionb{2.2}{The starburst galaxy M82}

\begin{figure}[t!]
\centering
\includegraphics[width=0.49\textwidth]{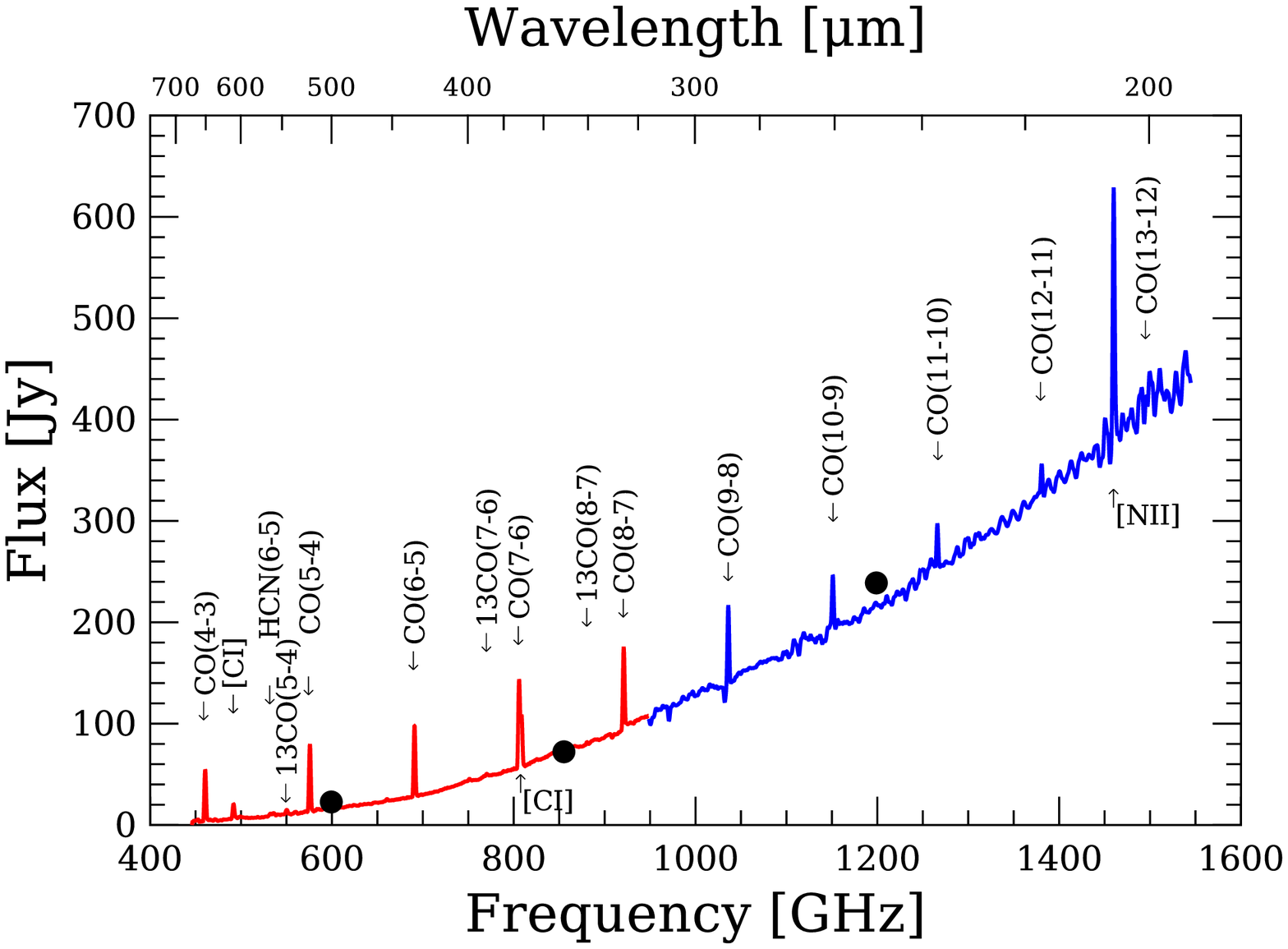}
\includegraphics[width=0.49\textwidth]{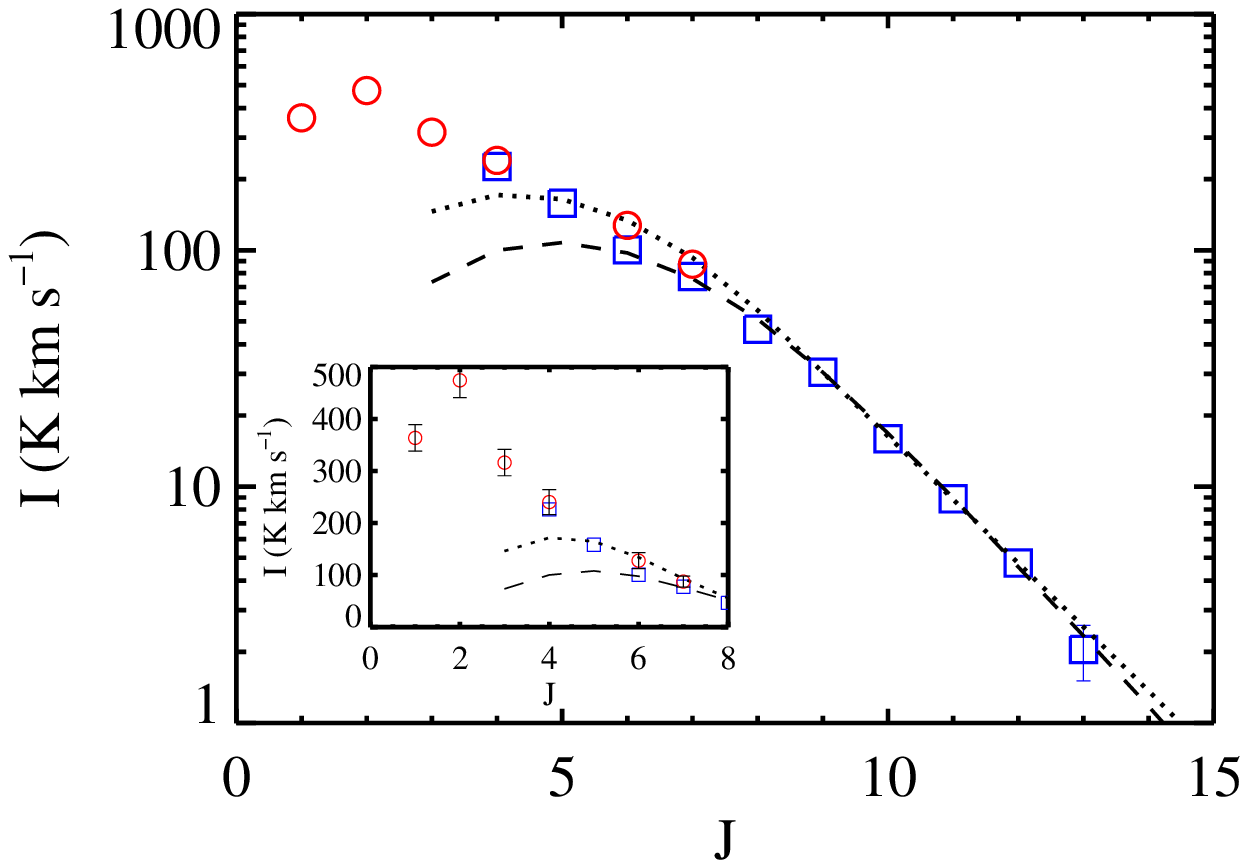}
\caption{{\bf{Left}}: Apodized spectrum of M82 corresponding to a 43.4
  arcsec beam, where red and blue lines represent data from the long-
  and the short-wavelength FTS bands respectively. Filled circles show
  SPIRE fluxes measured in the same beam. {\bf{Right}}: Comparison of
  the best fitting NLTE radiative transfer model (dotted line) with
  our CO line intensities. The model shown by a dashed line was
  obtained by using only $J\geq7-6$ CO lines. Open squares are the
  SPIRE FTS data, open circles are ground-based data from Ward et
  al.~(2003). The inset highlights the deviations from models at the
  lower $J$ end. }
\end{figure}

At a distance of 3.9 Mpc, M82 is the most well-studied starburst
galaxy in the local universe, and it is widely used as a starburst
prototype in cosmological studies. The ISM of this galaxy has been
mapped using ground-based observations, in particular the lowest-$J$
CO rotational lines that provide constraints on the physical state of
the cold molecular gas. With a molecular gas content of
$\sim1.3\times10^9~M_\odot$ (Walter et al. 2002), it served as a
perfect target for SPIRE FTS performance verification. M82 was
observed in the high spectral resolution point-source mode, on 2009
September 21; the total integration time was 1332 s.

Figure~1 (left) shows the entire SPIRE FTS spectrum of M82, which
displays a prominent CO emission-line ladder (from $J=4-3$ to
$J=13-12$) along with several molecular lines from HCN and $^{13}$CO,
and atomic fine-structure lines from [C{\sc{i}}] and [N{\sc{ii}}]. The
CO line intensities were modelled using the NLTE radiative transfer
code RADEX (van der Tak et al. 2007). We generated a grid of models
with different values in kinetic temperature, gas density and CO
column density. Our modelling (Figure 1, right) strongly indicates
that the observed CO emission is coming from very warm gas with a
total mass of $1.2\times10^7~M_\odot$ and a kinetic temperature of
540~K. Theoretical models predict that, at these temperatures, H$_2$
will be the dominant coolant compared to CO: in the case of M82, the
SPIRE detected molecular gas is expected to radiate about
$3\times10^7$~$L_\odot$ in H$_2$ lines. This is in good agreement with
values derived from ISO and Spitzer MIR spectroscopy (Rigopoulou et
al. 2002; Beir\~{a}o et al.\ 2008).

The heating source of this warm molecular gas is not easy to
uncover. A first possibility are UV-powered PDRs, but this can be
disregarded as the observed CO lines in M82 are far too luminous
compared to PDR models. Hard X-rays from an AGN have the potential for
heating molecular gas in an XDR, but there is no strong evidence for
an AGN in M82. Moreover, with a strong XDR component, such as seen in
Mrk\,231 (van der Werf et al.\ 2010), the spectral line energy
distribution becomes flat at high $J$ instead of decreasing as in
M82. Another scenario is heating via the enhanced cosmic ray density
generated by the high supernova rate in the nuclear starburst (Suchkov
et al.\ 1993), but the energy input per mass in M82 is too low to
match the observed cooling. The most probable heating mechanism is
turbulence from stellar winds and supernovae. A large velocity
gradient is required in order to reach the observed cooling rate of
2.6~$L_\odot/M_\odot$, but given the powerful stellar winds in the M82
starburst, this may not be unreasonable.

\subsectionb{2.3}{The ULIRG Arp\,220}

\begin{figure}[t!]
\centering
\includegraphics[width=\textwidth]{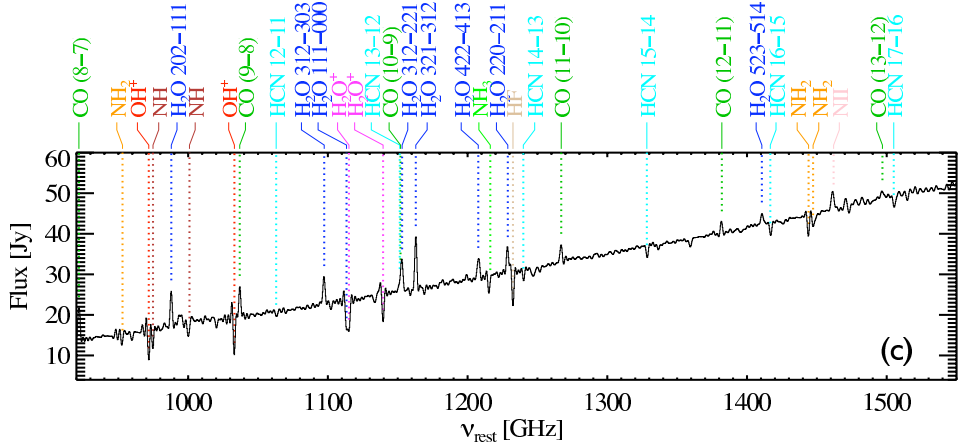}
\caption {The spectrum of Arp\,220 between 920 and 1550 GHz,
  corresponding to roughly half of the frequency range observed with
  SPIRE FTS. Different line identifications are indicated. }
\end{figure}
 
Arp\,220 is the nearest Ultra Luminous Infrared galaxy (ULIRG) at a
distance of about 77 Mpc. It is one of the most popular templates for
studies of high-$z$ dusty galaxies. It has two merging nuclei
separated by $\sim$1~arcsec and together they have a large reservoir
of molecular gas ($\sim10^{10}~M_\odot$). Arp\,220 has been observed
extensively over the years accross the electromagnetic spectrum. The
extreme star formation environment in this galaxy provides an
excellent laboratory to understand the processes affecting star
formation and possibly AGN feedback.

Arp\,220 was observed with the SPIRE FTS in the high spectral
resolution, single pointing mode on 2010 February 13. The total
on-source integration time was 10\,445 seconds. The spectrum
(Figure~2) shows the far-infrared (FIR) continuum and the detection of
several key molecular and atomic species. We detect a luminous CO
emission ladder from $J=4-3$ to $J=13-12$ and several water
transitions with total water luminosity comparable to CO. In addition
to the emission line features, the spectrum also shows several
absorption features. Most surprising are the detections of five very
high-$J$ HCN lines in absorption; their low-$J$ transitions are
detected in emission with ground-based observatories (Greve et al.\
2009). Such high-$J$ transitions of HCN have never been detected
before in external galaxies, in emission or absorption. Strong
absorption lines are also detected from hydrides, including three
transitions of OH$^+$, one of CH$^+$ and four of H$_2$O$^+$. The
detections of OH$^+$ and H$_2$O$^+$ are very important in Arp\,220 for
modeling the water transitions because they are the major
intermediaries in the ion-neutral chemistry network producing water in
the ISM, and their detection is very difficult from current
ground-based observatories.

We again used RADEX to interpret the CO spectral line energy
distribution. Our modelling strongly indicates that the mid-$J$ to
high-$J$ CO lines are tracing warm molecular gas with a kinetic
temperature of about 1350~K, whereas the low-$J$ transitions are
tracing cold gas at about 50~K. The mass of the warm gas is about 10\%
of the cold molecular gas but dominates the luminosity as well as the
cooling over the cold CO. The temperature of the warm molecular gas is
in excellent agreement with the temperature derived from H$_2$
rotational lines observed from Spitzer, implying that CO is still a
good tracer of H$_2$ at these high temperatures. Similar as for M82,
we can rule out PDRs, XDRs and cosmic rays as possible sources of this
warm molecular gas. The mechanical energy from supernovae and stellar
winds can satisfy the energy budget required to heat this gas but we
still do not know the exact mechanism that heats this gas. We also
used NLTE modelling to interpret the HCN spectral line energy
distribution. The transitions from emission to absorption takes places
somewhere between $J = 4-5$ and $J = 12-11$. These high-$J$ lines are
populated via infrared pumping of photons at 14 $\mu$m. The condition
for infrared pumping to populate $J = 17-16$ level requires an intense
radiation field with $T > 350$~K.

\begin{figure}[t!]
\centering
\includegraphics[width=0.32\textwidth]{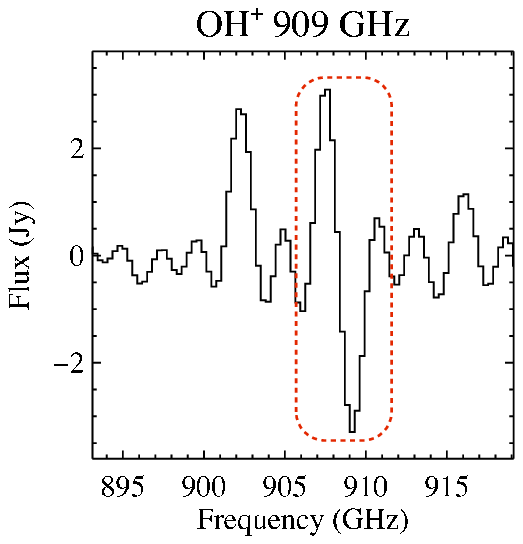}
\includegraphics[width=0.32\textwidth]{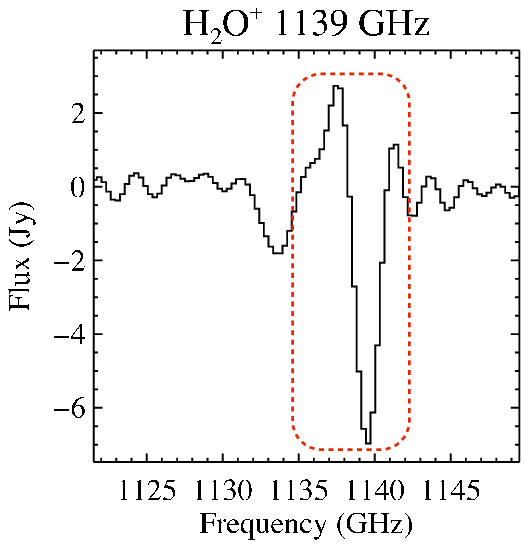}
\includegraphics[width=0.32\textwidth]{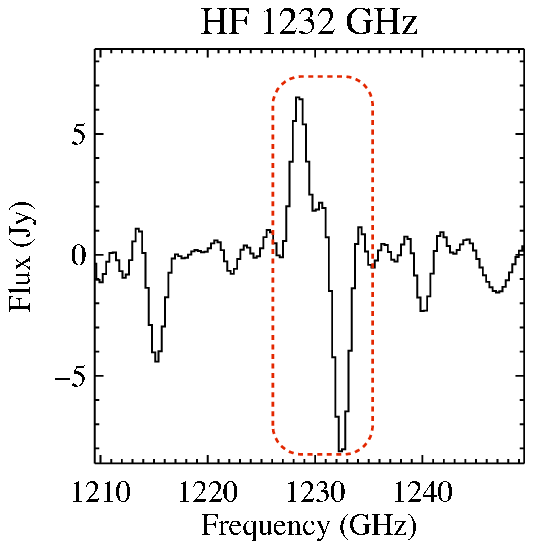}
\caption {P-Cygni profiles detected in OH$^+$, H$_2$O$^+$ and HF
  lines, suggesting molecular outflow in Arp\,220. }
\end{figure}
 
One of the most striking results from the SPIRE FTS data is the
detection of a massive molecular outflow, that could be driven by a
hidden AGN or starburst activity in Arp\,220.  Observations of massive
molecular outflows in galaxies can significantly improve our
understanding of the connection between AGN/starbursts-feedback and
galaxy evolution. It is believed that the energy injection from an AGN
or starbursts can quench star formation by expelling the molecular gas
out of the galaxy, transforming gas-rich blue galaxies to gas-poor red
galaxies. However, the evidence for massive molecular outflows had
been missing, until recently, when it was discovered in Mrk\,231 and
NGC\,1266.  The signature of a molecular outflow in Arp\,220 was seen
in the P-Cygni profiles of OH$^+$, H$_2$O$^+$ and H$_2$O
(Figure~3). Unfortunately our FTS spectra could not fully resolve this
outflow and we were only able to provide a lower limit on the mass of
the outflow ($\sim10^7~M_\odot$) and an upper limit on the velocity
($\sim250$ km\,s$^{-1}$). A high resolution follow-up with HIFI and
ALMA is required to accurately characterize this outflow.

\sectionb{3}{SPECTROSCOPY OF GRAVITATIONALLY LENSED HIGH-$z$ GALAXIES}

\subsectionb{3.1}{The H-ATLAS survey}

The Herschel ATLAS (H-ATLAS, Eales et al.\ 2010) is the largest
open-time key project that will be carried out on the Herschel Space
Observatory. It is surveying 570 deg$^2$ of the extragalactic sky, 4
times larger than all the other Herschel extragalactic surveys
combined, in five FIR/submm continuum bands. The main scientific goal
of the H-ATLAS is to provide measurements of the dust masses and
dust-obscured star formation for tens of thousands of nearby galaxies,
the FIR/submm equivalent to the SDSS photometric survey. However, the
H-ATLAS has many other science goals, ranging from studying cirrus and
debris discs in our own Milky Way (Thompson et al.\ 2010; Bracco et
al.\ 2011) and detailed studies of nearby galaxies (Baes et al.\ 2010)
to cosmological studies (Maddox et al.\ 2010; Clements et al.\ 2010).

One of the most fascinating results of H-ATLAS is the detection of a
significant number of gravitationally lensed objects in the distant
($z > 2$) Universe. This is possible in such a shallow but large area
survey because of the combined effects of a strong negative
K-correction and steep number counts in the SPIRE 500 $\mu$m
waveband. Objects brighter than 100 mJy at 500 $\mu$m will be a
mixture of lensed high-redshift galaxies, nearby galaxies and
flat-spectrum radio sources, and current models predict that H-ATLAS
will detect some 350 strongly lensed objects at 500 $\mu$m at $z>1$
(Negrello et al.\ 2007; Eales et al.\ 2010). High-$z$, bright submm
galaxies open a new possibility for the detailed study of the physical
conditions of the interstellar medium, at restframe FIR/submm
wavelengths: the important atomic emission lines shift towards submm
wavelengths where they can be observed with SPIRE FTS, whereas the
high-$J$ molecular lines, particularly the CO lines, shift towards
radio wavelengths, where they can be observed with ground-based radio
interferometers. The H-ATLAS consortium has set up a substantial
programme of follow-up observations of candidate gravitational lenses,
using both ground-based telescopes (IRAM, EVLA, SMA, GBT, Keck,\ldots)
and space missions (Herschel, Spitzer, HST).

\subsectionb{3.2}{The gravitationally lensed submm galaxy SDP.81}

\begin{figure}[t!]
\centering
\includegraphics[width=\textwidth]{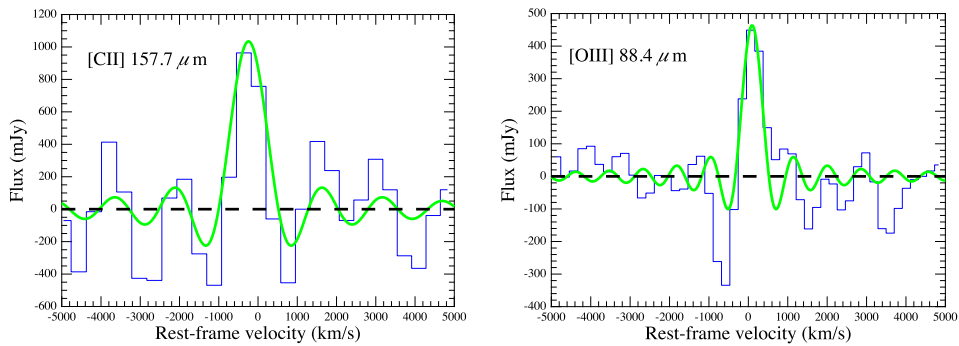}
\caption{Continuum subtracted regions around the [C{\sc{ii}}]
  157.7~$\mu$m (left) and [O{\sc{iii}}] 88.4~$\mu$m (right) lines of
  SDP.81. The SPIRE FTS data are shown as a histogram and the best fit
  sinc function is also shown as the solid line. }
\end{figure}
 
One of the best-studied H-ATLAS gravitational lenses so far is H-ATLAS
J090311.6+003906 (SDP.81), detected first in the H-ATLAS Science
Demonstration Phase field. A combination of radio and optical spectra
shows that this source is a submm galaxy at $z=3.04$, lensed by an
elliptical galaxy at $z=0.30$ (Negrello et al.\ 2010; Frayer et al.\
2011). SMA imaging reveals the submm morphology, consistent with a
lensing event, with multiple peaks distributed around the position of
the foreground elliptical galaxy (Negrello et al.\ 2010). Detailed
lens modelling based on multi-wavelength data, including deep Keck and
Spitzer imaging, revealed that the magnification factor is $25\pm7$,
and that the submm galaxy has a high dust obscuration ($A_V\sim4.4$)
and a star formation rate of $\sim75~M_\odot$\,yr$^{-1}$ (Hopwood et
al.\ 2011).

In order to study its ISM, we observed SDP.81 with the EVLA and SPIRE
FTS on 2010 July 17-18 and 2010 June 1, respectively (Valtchanov et
al.\ 2011). The most prominent features in the FTS spectrum of SDP.81
are the [O{\sc{iii}}] 88~$\mu$m and [C{\sc{ii}}] 158~$\mu$m lines
(Figure~4). The EVLA observations reveal a clear detection of the
CO(1-0) line, which is redshifted to 28.53~GHz, into the EVLA's new
Ka-band receivers. The line intensities of these lines, combined with
the continuum fluxes, were analysed using the PDR models of Kaufman et
al.\ (1999). The best fitting models indicate a PDR cloud-ensemble
density $n \approx 2000$~cm$^{-3}$ and a far-UV ionizing field
strength $G_0\sim200$. These characteristics are similar to other
high-$z$ star-forming galaxies. The radio observations provide
supporting evidence to the hints from the ISM lines, that part the
ionization field in the galaxy may be due to an AGN.

%\begin{figure}[!tH]
%\vbox{
%\centerline{\psfig{figure=fig1.eps,width=100mm,angle=0,clip=}}
%\vspace{1mm}
%\captionb{1}
%{Synthetic reddening line for the Kurucz model $T_{\rm eff}$ =
%4500 K, log\,$g$ = 4.0 (white circles) plotted on the observational
%2MASS $J$--$H$ vs.~$H$--$K_s$ diagram in the direction with the Galactic
%coordinates $\ell$ = 330\degr, $b$ = 0\degr.}
%}
%\end{figure}

\sectionb{4}{SUMMARY AND CONCLUSIONS}

The FIR/submm domain is a fascinating region to study the physics of
the ISM, as it contains a large variety of tracers of both dust
emission and the different gas phases. The launch of the Herschel
Space Observatory has finally opened up this region for systematic
spectroscopic studies. We have presented a number of new spectroscopic
results obtained with Herschel in the frame of the VNGS and H-ATLAS
key programmes. Concerning nearby galaxies, we have presented, based
on an impressive CO ladder, the detection of warm and dense molecular
gas in M82 and Arp\,220, probably heated by mechanical energy from the
starburst. Moreover, in Arp\,220, we have detected many molecular
species (including hybrides) which support the evidence for a hidden
AGN, and we have interpreted the observed P-Cygni profiles in various
molecular lines as a massive molecular outflow. Concerning distant
submm galaxies, we have discussed the case of SDP.81, a submm galaxy
at $z=3.04$ lensed by a foreground elliptical with a magnification of
about 25. We have presented SPIRE FTS and radio spectroscopy of this
lensed submm galaxy and demonstrated that these measurements are
unique probes of the ISM of high-$z$ galaxies. The results we
presented are just the first results obtained in the frame of the VNGS
and H-ATLAS surveys, and many more results (both detailed studies on
individual objects and more statistical studies on larger samples) are
expected to come out in the next few years.

% \thanks{ The authors are thankful to A and B for their help. The use of
% the Simbad, 2MASS and SkyView databases is acknowledged.}

\References

\refb Auld R., Smith M.~W.~L., Bendo G., et al., 2011, MNRAS, submitted

\refb Baes M., Fritz J., Gadotti D.~A., et al., 2010, A\&A, 518, L39

\refb Beir\~{a}o P., Brandl B.~R., Appleton P.~N., et al., 2008, ApJ, 676, 304

\refb Bendo G.~J., Wilson C.~D., Pohlen M., et al., 2010, A\&A, 518, L65

\refb Bendo G.~J., Boselli A., Dariush A., et al., 2011, MNRAS, submitted

\refb Bracco A., Cooray A., Veneziani M., et al., 2011, MNRAS, 412, 1151

\refb Brauher J.~R., Dale D.~A., Helou G., 2008, ApJS, 178, 280

\refb Clements D.~L., Rigby E., Maddox S., et al., 2010, A\&A, 518, L8

\refb Davies J.~I., Wilson C.~D., Auld R., et al., 2010, MNRAS, 409, 102

\refb de Graauw Th., Helmich F.~P., Phillips T.~G., et al., 2010, A\&A, 518, L6

\refb Eales S., Dunne L., Clements D., et al., 2010, PASP, 122, 499

\refb Frayer D.~T., Harris A.~I., Baker A.~J., et al., 2011, ApJ, 726, L22

\refb Greve T.~R., Papadopoulos P.~P., Gao Y., Radford S.~J.~E., 2009,
ApJ, 692, 1432

\refb Griffin M.~J., Abergel A., Abreu A., et al., 2010, A\&A, 518, L3

\refb Hopwood R., Wardlow J., Cooray A., et al., 2011, ApJ, 728, L4

\refb Kaufman M.~J., Wolfire M.~G., Hollenbach D.~J., Luhman M.~L., 1999, ApJ, 527, 795

\refb Maddox S.~J., Dunne L., Rigby E., et al., 2010, A\&A, 518, L11

\refb Negrello M., Perrotta F., Gonz\'alez-Nuevo J., et al., 2007, MNRAS, 377, 1557 

\refb Negrello M., Hopwood R., De Zotti G., et al., 2010, Science, 330, 800

\refb Panuzzo P., Rangwala N., Rykala A., et al., 2010, A\&A, 518, L37

\refb Pilbratt G.~L., Riedinger J.~R., Passvogel T., et al., 2010, A\&A, 518, L1

\refb Poglitsch A., Waelkens C., Geis N., et al., 2010, A\&A, 518, L2

\refb Rangwala N., Maloney P.~R., Glenn J., et al., 2011, MNRAS,
submitted (arXiv:1106.5054)

\refb Rigopoulou D.,Kunze D., Lutz D., et al., 2002, A\&A, 389, 374

\refb Roussel H., Wilson C.~D., Vigroux L., et al., 2010, A\&A, 518, L66

\refb Spaans M., Meijerink R., 2008, ApJ, 678, L5 

\refb Suchkov A., Allen R.~J., Heckman T.~M., 1993, ApJ, 413, 542

\refb Thompson M.~A., Smith D.~J.~B., Stevens J.~A., et al., 2010, A\&A, 518, L134

\refb Tielens A.~G.~G.~M., Hollenbach D., 1985, ApJ, 291, 722

\refb Valtchanov I., Virdee J., Ivison R.~J., et al., 2011, MNRAS, in press (arXiv:1105.3924)

\refb van der Tak F.~F.~S., Black J.~H., Sch\"oier F.~L., et al.\ 2007, A\&A, 468, 627

\refb van der Werf P.~P., Isaak K.~G., Meijerink R., et al.\ 2010, A\&A, 518, L42

\refb Walter F., Weiss A., Scoville N., 2002, ApJ, 580, L21

\refb Ward J.~S., Zmuidzinas J., Harris A.~I., Isaak K.~G., 2003, ApJ, 587, 171

\end{document}